\documentclass[
reprint,
superscriptaddress,
amsmath,amssymb,
aps,
pra,
10pt]{revtex4-2}

\usepackage{graphicx, subfigure}
\usepackage{dcolumn}
\usepackage{bm}
\usepackage{hyperref}
\hypersetup{
    colorlinks=true,
    citecolor=blue,
    linkcolor=black,
    filecolor=magenta,      
    urlcolor=ForestGreen,
    pdfpagemode=FullScreen,
    }
\usepackage{physics}
\usepackage{dsfont}
\usepackage{amsmath}
\usepackage{braket}
\usepackage{amsfonts}
\usepackage{leftindex}
\usepackage{amssymb}
\usepackage{siunitx}
\usepackage[dvipsnames]{xcolor}
\usepackage{soul}
\usepackage{multirow}
\sethlcolor{Yellow}
\usepackage{todonotes}

\begin{document}
	
 	\title{Time-resolved and Superradiantly Amplified Unruh Effect}

    \author{Akhil Deswal}
	\email{akhildeswal.phy@gmail.com}
	\affiliation{Department of Physical Sciences, Indian Institute of Science Education \& Research (IISER) Mohali, Sector 81 SAS Nagar, Manauli PO 140306 Punjab India.}
 
	\author{Navdeep Arya}
    \email{navdeep.arya@fysik.su.se}
    \email{navdeeparya.me@gmail.com}
    \affiliation{Department of Physics, Stockholm University, Roslagstullsbacken 21, 106 91 Stockholm, Sweden}

    \author{Kinjalk Lochan}
	\affiliation{Department of Physical Sciences, Indian Institute of Science Education \& Research (IISER) Mohali, Sector 81 SAS Nagar, Manauli PO 140306 Punjab India.}
	
	\author{Sandeep K.~Goyal}
	\affiliation{Department of Physical Sciences, Indian Institute of Science Education \& Research (IISER) Mohali, Sector 81 SAS Nagar, Manauli PO 140306 Punjab India.} 
	

\begin{abstract}
We identify low-acceleration conditions under which the Unruh effect manifests as an early superradiant burst in a collection of excited atoms. The resulting amplified Unruh signal is resolved from the inertial signal both in time and intensity. We demonstrate theoretically that these conditions are realized inside a sub-resonant cavity that highly suppresses the response of an inertial atom, while allowing significant response from an accelerated atom as, owing to the acceleration-induced spectral broadening, it can still couple to the available field modes. The setup thus selectively amplifies the modified field fluctuations underlying the Unruh effect into an early superradiant burst. In comparison, the field fluctuations perceived inertially would cause a superradiant burst much later. In this way, we simultaneously address the extreme acceleration requirement, the weak Unruh signal, and the dominance of the inertial signal, all within a single experimental arrangement. 
\end{abstract}
	\maketitle
 
\paragraph*{Introduction\textemdash} The quantum field fluctuations are known to be altered under {various} conditions, leading to phenomena like the Casimir effect~\cite{Casimir1948}, the Schwinger effect~\cite{Schwinger1951}, Hawking radiation~\cite{Hawking1974}, and particle creation in an expanding universe~\cite{Parker1968}. Additionally, the concepts of vacuum and particle content of a quantum field are inherently observer-dependent, as elegantly encapsulated in the Fulling-Davies-Unruh effect~\cite{Fulling1973,Davies1975,Unruh1976,Lima2019}\textemdash predicting that a uniformly accelerated observer perceives the inertial vacuum of a free quantum field to be in a thermal state at a temperature $T_{\rm U}$ proportional to its acceleration $a$: $T_{\rm U} = \hbar a/2 \pi k_{\rm B} c$. 

While individual atoms provide valuable insights into such phenomena, a collection of atoms can function as an even more sophisticated probe~\cite{Arya2024b}, leveraging the rich dynamics of collective effects~\cite{Dicke1954,Gross1982}.
For instance, photon emission from an extended sample of $N$ excited atoms sensitively depends on the distribution of atoms in the sample and the properties of the electromagnetic field to which the atoms are coupled~\cite{Eberly1969,Eberly1971,Eberly1972,Agarwal1970a}. Although a single excited atom decays spontaneously with an exponential profile at an emission rate $\gamma$, the collective behavior of a group of atoms can significantly modify the emission process~\cite{Dicke1954}. In an array of excited atoms where the atoms are indistinguishable with respect to their coupling to the electromagnetic field, correlations begin to develop between them as soon as any atom in the array undergoes spontaneous emission~\cite{Gross1982}. The correlations build up during a time period $0<\tau<\tau_{\rm d}$, ultimately leading to an intense event of photon emission, known as the superradiant burst, at the \textit{superradiant delay time} $\tau_{\rm d}$. The delay time is thus sensitive to the field fluctuations experienced by the atomic sample, and therefore should respond to, for example, the sample's acceleration or the presence of a gravitational field~\cite{Arya2024b}.  The intense photon emission rate lasts for a short period $\tau_{\rm sr}$, known as the superradiance time. The superradiance process features a maximum emission rate scaling super-linearly with the total number of participating atoms, well-defined directionality, and a time-resolving nature characterized by the superradiant delay time $\tau_{\rm d}$ and the superradiance time $\tau_{\rm sr}$~\cite{Dicke1954,Gross1982}.

\begin{figure}
    \centering
    \includegraphics[height=5.5cm]{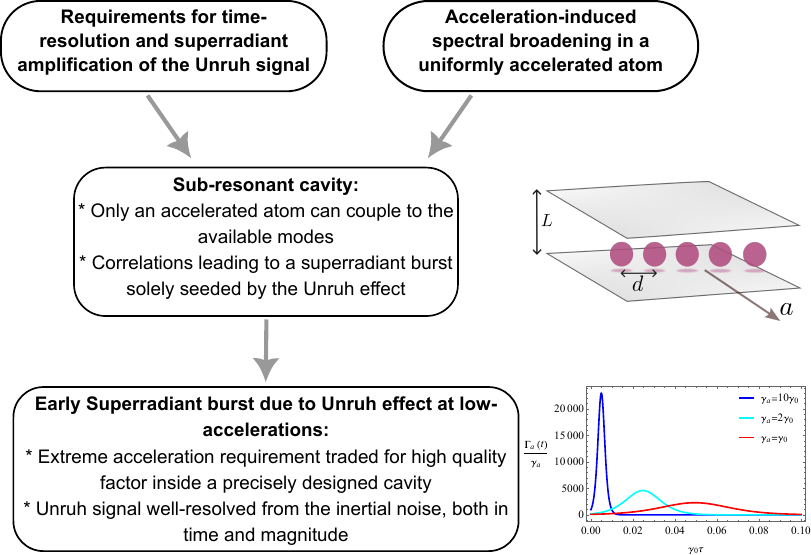}
    \caption{Graphical summary of the results.}
    \label{fig:summary}
\end{figure}

The Unruh effect, of great interest partly due to its close connection to the Hawking effect~\cite{Mukhanov2007}, remains untested due to the requirement of extreme accelerations. At achievable accelerations, the expected signal, being extremely weak, is easily overwhelmed by the inertial noise and the noise due to ambient laboratory temperature. 
In this Letter, we leverage the acceleration-induced non-resonant behavior of an atom, inside a sub-resonant cavity, in combination with the hallmarks of superradiance to obtain a time-resolved and highly amplified Unruh signal at low accelerations. Refer to Fig.~\ref{fig:summary} for a graphical summary of the results. 
\paragraph*{Collective response of atoms\textemdash} 
Consider $N$ identical two-level atoms forming a one-dimensional ordered array with interatomic spacing $\hat{y} d$, transverse to their motion along $\hat{z}$-direction. The atoms are coupled to electromagnetic field between two parallel mirrors with separation $\hat{x} L$ and reflectivity $R$. Each atom has a transition frequency $\omega_0$ and the corresponding transition wavelength is denoted by $\lambda_0$. We address an array subjected to uniform linear acceleration, in which each atom takes the trajectory $t(\tau) = a^{-1} \sinh(a \tau),~ z(\tau) = a^{-1} \cosh(a \tau)$, as a Rindler array. Here, $\tau$ is the proper time of an atom.
The total photon emission rate of such an array is obtained as~[see Supplemental Material~\cite{SM} for details]
\begin{equation}\label{eq:tdr}
		\Gamma_{\rm a}(\tau) = \frac{\gamma_{\rm a}}{4 \mu_{\rm a}}(\mu_{\rm a} N +1)^2  \sech^2 \Bigg( \frac{\tau - \tau_{\rm d}}{2 \tau_{\rm sr}}  \Bigg),
\end{equation}	
where $\gamma_{\rm a}$ is the emission rate of a single Rindler atom, $ \tau_{\rm d} \equiv \ln(\mu_{\rm a} N)/\gamma_{\rm a}(\mu_{\rm a} N+1 ) $ is the superradiant delay time, and $ \tau_{\rm sr} \equiv 1/\gamma_{\rm a}(\mu_{\rm a} N+1) $ is the superradiance time as marked in Fig.~\ref{fig:col-vs-inc}. 
The \textit{shape factor} $\mu_{\rm a}$ of the array is defined as $\mu_{\rm a} \equiv \gamma^{-1}_{\rm a} N^{-2} \sum_{i\ne j} \gamma^{\rm (a)}_{ij}$; where $\gamma^{\rm (a)}_{ii}=\gamma_{\rm a}$ and for $ i\ne j $, $\gamma^{\rm (a)}_{ij}$ quantifies the extent to which $i$th and $j$th atoms in the array influence each other's dynamics.
The shape factor depends on the distribution of atoms in the sample and the properties, like mode structure and particle content, of the field to which the atoms are coupled. In addition, it is sensitive to the acceleration of the array and the presence of any gravitational effects~\cite{Arya2024b}. 
\begin{figure}
    \centering
    \includegraphics[width=0.9\linewidth]{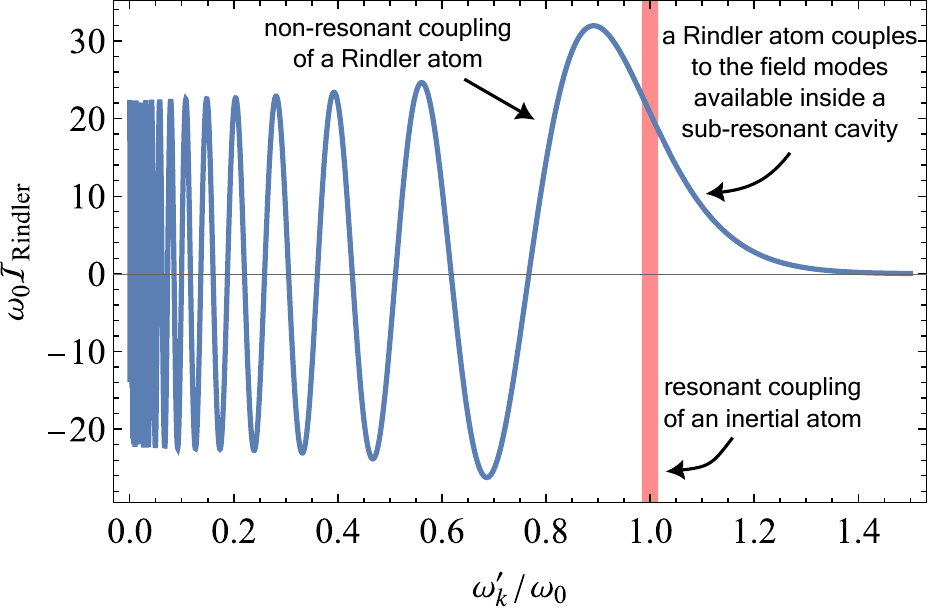}
    \caption{\textbf{Acceleration-induced spectral broadening in a Rindler atom.} $\mathcal{I}_{\text{Rindler}}$ is the spectral response of a uniformly accelerated atom. The red band shows an inertial atom's resonant coupling, i.e., to modes with $\omega'_k \approx \omega_0$. The blue curve shows non-resonant coupling of a Rindler atom to field modes. The plot is for $a/\omega_0 = 10^{-1}$.}
    \label{fig:AISB1}
\end{figure}
\begin{figure*}
\centering
  \subfigure[]{
	\includegraphics[height=3.6cm]{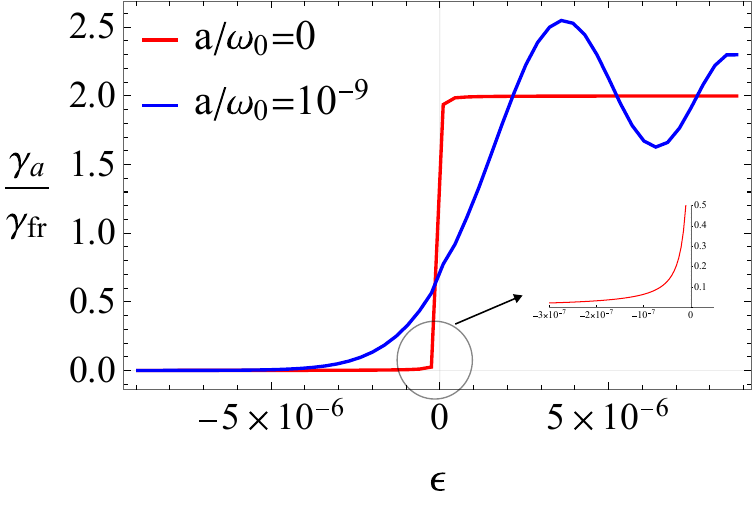} 
  \label{fig:nigamma}}
  \subfigure[]{
	\includegraphics[height=3.6cm]{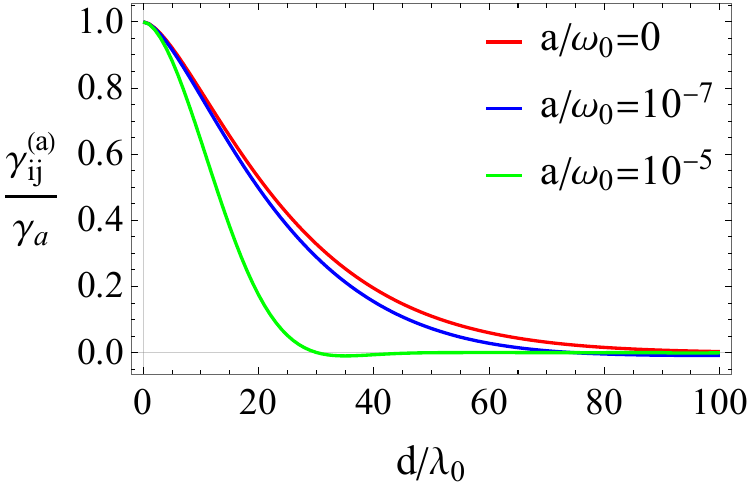}
    \label{fig:gammaij1}}
  \subfigure[]{
    \includegraphics[height=3.6cm]{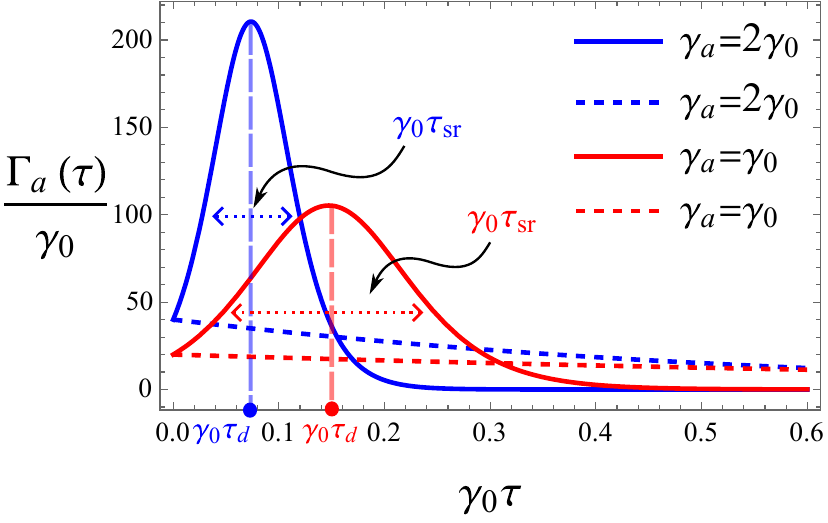}
    \label{fig:col-vs-inc}}
\caption{\textbf{Interplay of acceleration-induced spectral broadening and collective effects inside a sub-resonant cavity.} (a) Behavior of the emission rate of an inertial and a Rindler atom as a function of the cavity detuning parameter $\epsilon$ (defined as $\omega_0 L = \pi + \epsilon$), for $R=1-10^{-8}$. Here, $\gamma_{\rm fr}$ is the spontaneous emission rate of a single inertial atom in free space. For $\epsilon < 0$, that is $L < \lambda_0/2$, the emission rate of an inertial atom is highly suppressed, whereas the emission rate of a Rindler atom falls to the same extent for much lower mirror separation. (b) Impact of acceleration on $\gamma^{\rm (a)}_{ij}/\gamma_{\rm a}$, the cooperation between $i$th and $j$th atoms, as a function of separation between the two atoms inside a sub-resonant cavity with $R = 1 - 10^{-4}$ and $\omega_0 L = \pi-10^{-4}$. For a given separation between the two atoms, cooperation between them diminishes for increasing acceleration. (c) Comparison of the temporal behavior of the emission rate of an incoherent sample (dashed curves) of atoms versus that of a superradiant sample (solid curves), for $N=20$. The temporal emission profiles of inertial and Rindler samples of independent atoms considerably overlap with each other in time. In superradiant samples however, the two signals can be resolved temporally as the superradiance process has a time-resolving nature characterized by the delay time $\tau_{\rm d}$ and the superradiance time $\tau_{\rm sr}$.} 
 \label{fig:setup}
\end{figure*}
\paragraph*{Condition for time-resolution and superradiant enhancement of the Unruh signal\textemdash}
An intuitive way to appreciate the shape factor is to think of $\mu N$ as the \textit{effective number of cooperating atoms}~\cite{Eberly1971}. In the small sample limit, the so-called Dicke regime~\cite{Dicke1954}, $\mu N \to N - 1$, meaning that all the atoms in the sample cooperate, that is, the cooperative effects are the strongest. In general, under the effect of uniform acceleration, $\gamma_{\rm a}/\gamma_0$ increases while $\mu_{\rm a}/\mu_{0}$ decreases as a function of acceleration (see Fig.~\ref{fig:gammaij1}). We are interested in resolving the superradiant burst of a Rindler array from that of an inertial array. The occurrence of the two superradiant peaks will differ if the corresponding superradiant delay times are different enough. Further, the overlap of the two superradiant temporal profiles can be reduced if the collective effects are not compromised much due to the acceleration\textemdash this ensures that $\tau_{\rm sr}$, the time period over which superradiance occurs, remains small. 

To this end, consider the ratio of the superradiant delay times for a Rindler and an inertial array:
\begin{equation}\label{cdn1}
    \frac{\tau^{\rm (a)}_{\rm d}}{\tau^{(0)}_{\rm d}} = \frac{\gamma_0 (\mu_0 N + 1) \ln(\mu_{\rm a} N)}{\gamma_{\rm a}(\mu_{\rm a} N + 1)\ln(\mu_{0} N)}.
\end{equation}
The two signals are well-resolved in time if $\tau^{\rm (a)}_{\rm d} < \tau^{\rm (0)}_{\rm d}$.
If we chose $d/\lambda_0$ such that both $\mu_{\rm a} N$ and $\mu_{\rm 0} N$ are much greater than $1$ (ensuring strong collective effects), the requirement simplifies to
\begin{equation*}\label{cdn2}
    \frac{\tau^{\rm (a)}_{\rm d}}{\tau^{(0)}_{\rm d}} \approx \frac{\gamma_0 \mu_0 N }{\gamma_{\rm a}\mu_{\rm a} N } \frac{\ln(\mu_{\rm a} N)}{\ln(\mu_{0} N)} < 1.
\end{equation*}
At the same time, since the maximum superradiant emission rate scales as $(\mu N)^2$, if we want time-resolution of the two signals without compromising on superradiant amplification of the noninertial signal, we additionally require $\mu_{\rm a}/\mu_0 \approx 1$ (with $\mu_{\rm a}$ and $\mu_{0}$ each individually close to unity). Thus, the requirements to \textit{time-resolve and superradiantly amplify} the Unruh signal are $\gamma_{\rm a}/\gamma_0 \gg 1$ and $\mu_{\rm a}/\mu_0 \approx 1$. Under these conditions, the correlations between atoms that eventually lead to a superradiant burst build much faster in a Rindler array, with the buildup exclusively driven by the Unruh effect. The superradiant burst would thus be solely seeded by the modified field fluctuations underlying the Unruh effect predicted to be experienced by an accelerated system. This early superradiant burst would serve as a clear signature of the Unruh effect. It is noteworthy that the inertial vacuum state restricted to either the left or right Rindler wedge is in fact a thermal state at temperature $T_{\rm U}$ in the kind of cavity setup under consideration here---that is, where the cavity modifies the density of field modes only in the direction transverse to the observer's acceleration~\cite{Arya2024ULS}. Since $\tau^{\rm (a)}_{\rm d}/\tau^{(0)}_{\rm d} \approx \gamma_0/\gamma_{\rm a}$ under the conditions stated above, delay-time measurements in a setup first benchmarked with an inertial sample of atoms (i.e. $\gamma_0$, $\mu_0$, and $t^{(0)}_{\rm d}$ determined) can be used to obtain $\gamma_{\rm a}$. These values can then be verified against the theoretical prediction for $\gamma_{\rm a}$. Consequent upon this verification, one may infer the corresponding Unruh temperature directly from the expression $T_{\rm U} = \hbar a/2 \pi k_{\rm B} c$.

To understand the effect of acceleration on $\mu$, in Fig.~\ref{fig:gammaij1} we analyze the behavior of $\gamma^{\rm (a)}_{ij}/\gamma_{\rm a}$ as a function of $d/\lambda_0$ for different accelerations. Note that for a larger acceleration $(a/\omega_0 \sim 10^{-5})$, the cooperation between $i$th and $j$th atoms reduces quicker with increasing $d/\lambda_0$. On the other hand, for a smaller acceleration $(a/\omega_0 \sim 10^{-7})$, the cooperation is almost the same as for the inertial case over the range of $d/\lambda_0$ shown. Therefore, the larger the acceleration, quicker is the fall in $\gamma^{\rm (a)}_{ij}/\gamma_{\rm a}$ with increasing $d/\lambda_0$. Importantly, a finite, yet small, interatomic distance is required to preserve collective effects against dephasing due to coherent dipole-dipole interactions~\cite{Gross1982}. This effect in our setup is analyzed in the Supplemental Material~\cite{SM}. 
Therefore, the requirement of $\mu_{\rm a}/\mu_0 \approx 1$ can be comfortably fulfilled at lower accelerations.

Further, note that in general the emission rate $\gamma_{\rm a}$ of an accelerated atom can be written as $\gamma_{\rm a} = \gamma_0 + \tilde{\gamma}(\alpha)$, $\alpha \equiv a/\omega_0$, where the last contribution is purely-noninertial, that is, $\lim_{\alpha \to 0} \tilde{\gamma}(\alpha) = 0$. As we have already noted that $\mu_{\rm a}/\mu_0 \approx 1$ can be achieved at low accelerations, clearly, the time-resolution and superradiant enhancement of the Unruh signal hinges on achieving $\gamma_{\rm a}/\gamma_0 \gg 1$. For $\gamma_{\rm a}/\gamma_0 = 1 + \tilde{\gamma}/\gamma_0$ to be much greater than unity, we require $\tilde{\gamma}/\gamma_0 \gg 1$, which could not be achieved in any of the proposals for observing the Unruh effect so far.

\begin{figure*}
	\centering
	\subfigure[]{
		\label{fig:ratiogamma123}
		\includegraphics[height=3.6cm]{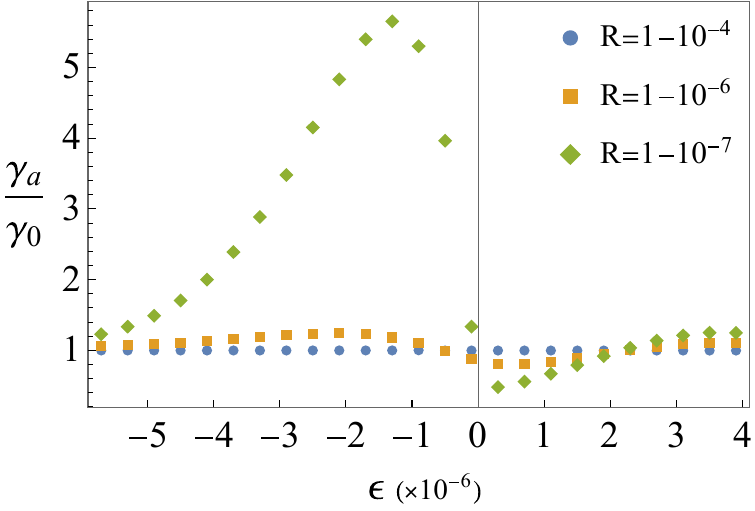} }
	\subfigure[]{
		\label{fig:ratiogamma}
		\includegraphics[height=3.6cm]{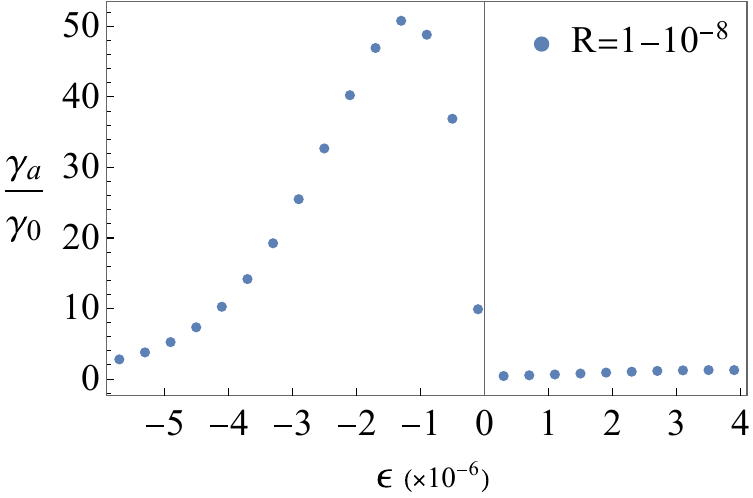} }
	\subfigure[]{
		\label{fig:ratiomu}
		\includegraphics[height=3.6cm]{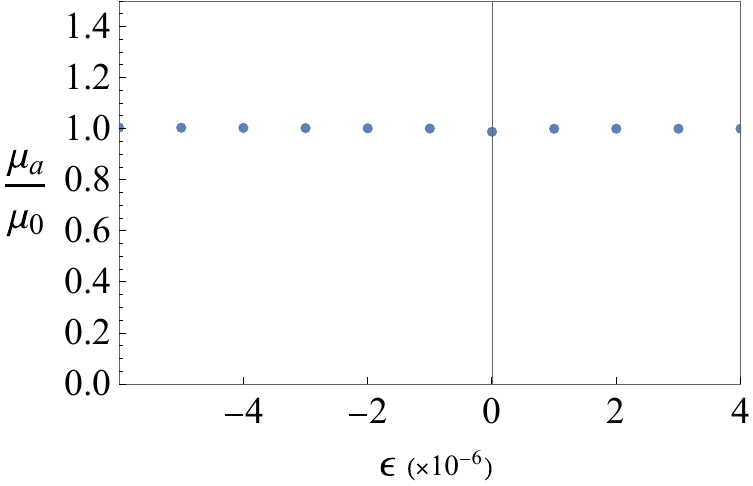} 
	}
	\caption{\textbf{Realization of conditions for time-resolution and superradiant enhancement of the Unruh signal inside a sub-resonant cavity.} (a),(b) Dependence of $\gamma_{\rm a}/\gamma_0$ on the reflectively (equivalently, quality factor) and the cavity detuning parameter $\epsilon$. Inside a sub-resonant cavity, higher mirror reflectivity leads to higher $\gamma_{\rm a}/\gamma_0$ values due to a stronger suppression of the emission rate of an inertial atom, while a Rindler atom still responds significantly due to the acceleration-induced non-resonant behavior. As $\gamma_{\rm a}/\gamma_0 = 1 + \tilde{\gamma}(\alpha)/\gamma_0$, unambiguously resolving the purely-noninertial signal $\tilde{\gamma}(\alpha)$ against the inertial signal $\gamma_0$ requires $\tilde{\gamma}(\alpha)/\gamma_0 \geq 1$, that is, $\gamma_{\rm a}/\gamma_0 \geq 2$. For $a/\omega_0 = 10^{-9}$, the two signals are not resolved unless the mirror reflectivity is equal to or better than $1-10^{-7}$, for which the two signals are well-resolved in a sub-resonant cavity configuration. This tradeoff between acceleration requirement and cavity's quality factor is elaborated in Table\,\ref{table}. In (b), the required precision in the specification of cavity width to access $\gamma_{\rm a}/\gamma_0 \gg 1$ is $\Delta L/L \sim 10^{-6}$. (c) The ratio $\mu_{\rm a}/\mu_0$ of the Rindler and inertial shape factors, for an atom array with $d/\lambda_0=1$, remains nearly constant over the cavity detuning range of interest. For all the plots, $a/\omega_0 = 10^{-9}$.}
	\label{fig:main}
\end{figure*}

Next, we demonstrate that $\tilde{\gamma}(\alpha)/\gamma_0 \gg 1$ can be achieved inside a sub-resonant cavity by optimally harnessing the acceleration-induced non-resonant behavior of a Rindler atom.
\paragraph*{Acceleration-induced spectral broadening\textemdash}
The acceleration induces a non-resonant behavior in an atom, broadening its spectrum. In Fig.~\ref{fig:AISB1}, we show the field modes that participate in the response of a Rindler atom, where $\omega_0 \mathcal{I}_{\rm Rindler}$ is the strength of participation of a given mode~(see Supplemental Material~\cite{SM} for more details). Note some intriguing features in a Rindler atom's coupling to the field modes: for mode frequencies $\omega_k' < \omega_0$, the atom couples constructively ($\mathcal{I}_{\rm Rindler} > 0$) to some field modes and destructively ($\mathcal{I}_{\rm Rindler} < 0 $) to others. More importantly, it couples (constructively) to modes having $\omega_k' > \omega_0$, a feature that can be optimally exploited inside a sub-resonant cavity wherein all the available field modes have frequency greater than the atom's transition frequency.
Earlier proposals~\cite{Jaffino2022,Arya2024ULS,Barman2025} using density of field modes to relax acceleration requirement did not fully exploit the acceleration-induced spectral broadening as they focused on cavities tuned above the first resonance point, limiting the signal-to-noise ratio $(\tilde{\gamma}/\gamma_0 \lesssim 1)$ that could be achieved in such setups, as analyzed in Figs.~\ref{fig:ratiogamma123},\ref{fig:ratiogamma}.
\paragraph*{Response inside a sub-resonant cavity\textemdash}
The spontaneous emission rate of a single inertial atom placed between two perfect mirrors ($R \to 1$) is given as $\gamma_0 = \frac{1}{L} \sum_{n=1}^{\infty} \sin[2](n\pi/2)   \Theta\left(1 - n \lambda_0/2 L\right)$~\cite{SM},
where $\Theta(x)$ is the Heaviside theta function.
Note that $\gamma_0$ receives contributions from field modes with $n < 2L/\lambda_0$. In particular, if the separation between the mirrors is less than $\lambda_0/2$, there are no field modes available to facilitate spontaneous emission from the atom, leading to a vanishing $\gamma_0$~\cite{Heinzen1987, Martini1987,Anderson1987}.
However, any realistic mirrors would have a reflectivity less than unity and therefore the emission rate for $L < \lambda_0/2$ decreases smoothly with decreasing $L$~\cite{SM}. 

On the other hand, due to acceleration-induced non-resonant behavior, a Rindler atom shows a stronger emission rate as the mirror separation is lowered below the first resonance point. Figure~\ref{fig:nigamma} compares the emission rates of an inertial and Rindler atom as a function of mirror separation and clearly shows a stronger emission rate of the Rindler atom below the first resonance point.
As a result, in the sub-resonant configuration of the cavity, a large $\gamma_{\rm a}/\gamma_0$ ratio can be obtained. However, the emission rate for even an accelerated single atom in this cavity configuration is extremely weak. But, with many cooperatively behaving atoms ($\mu_{\rm a}N \approx \mu_{0} N \gg 1$), which is possible at low accelerations with $d/\lambda_0$ lying in an appropriate range, the two rates can be superradiantly amplified. In flat spacetime, smaller interatomic spacing generally enhances the collective emission rate, provided that dephasing caused by coherent dipole-dipole interactions does not suppress the collective behavior~\cite{Gross1982}. The detrimental effects of these interactions can be mitigated by arranging the atoms in an orderly fashion such that each atom experiences a similar environment~\cite{Friedberg1974,Coffey1978,Gross1982}. For example, ordered arrays~\cite{Masson2020,Masson2022} or ring configuration~\cite{Friedberg1974,Coffey1978} would be suitable choices.

In Figs.~\ref{fig:ratiogamma123} and \ref{fig:ratiogamma}, we plot the ratio $\gamma_{\rm a}/\gamma_0$ as a function of the mirror separation for different values of mirror reflectivity. For reflectivities $R < 1$, mirror separations $L>\lambda_0/2$ lead to $\gamma_{\rm a}/\gamma_0 \sim 1$, that is, $\tilde{\gamma}/\gamma_0 \ll 1$. However, for mirror separations $L < \lambda_0/2$ and for a given acceleration of the Rindler atom, the ratio $\gamma_{\rm a}/\gamma_0$ increases with higher mirror reflectivities, taking values as large as $50$ for $R = 1 - 10^{-8}$, $a \sim 10^{-9} \omega_0 c$ (where we have momentarily restored $c$). Thus, the sub-resonant cavity configuration achieves $\tilde{\gamma}(\alpha)/\gamma_0 \gg 1$, as required. Moreover, in the same parameter range, $\mu_{\rm a}/\mu_{0} \approx 1$ as shown in Fig.~\ref{fig:ratiomu}. 

To summarize, combined with $\mu N \gg 1$, these conditions mean that the buildup of correlations among atoms, required for a superradiant burst, occurs faster in a Rindler array and is driven predominantly by the Unruh effect, as $\tilde{\gamma}(\alpha)/\gamma_0 \gg 1$. The field fluctuations perceived inertially under these conditions will be entirely inadequate to cause a superradiant burst in the given time. The early superradiant burst is thus entirely seeded by the modified field fluctuations underlying the Unruh effect experienced by the accelerated array of atoms\textemdash giving us a time-resolved and superradiantly amplified Unruh signal. A noticeable shift in $\tau_{\rm d}$ due to the Unruh effect is easier to isolate experimentally than a shift in intensity only. Due to overlapping emission timescales, a collection of independent atoms cannot provide such a temporal resolution. Also, note that modeling superradiance merely as a perfect amplifier introduces ambiguities regarding the cause of the superradiant burst. Due to the tradeoff between acceleration requirement and the cavity quality factor (Table\,\ref{table}), a burst seeded by the inertially-perceived vacuum fluctuations can very easily be conflated with the one caused by the Unruh effect. Therefore, as discussed in the Supplemental Materials~\cite{SM}, to employ the time-resolving nature of the superradiance process one must account for factors to which $\tau_{\rm d}$ and $\tau_{\rm sr}$ are intrinsically sensitive, these include cavity's finite quality factor, separation between the atoms, and dephasing due to coherent dipole-dipole interactions.
	\begin{table}
    \caption{\label{table} \textbf{Requirement of extreme accelerations can be traded for high quality factor inside a precisely designed cavity.} Minimum required quality factor, $Q_{\rm min} = 2 \pi L \sqrt{R_{\rm min}}/\lambda_0 (1 - R_{\rm min})$\,\cite{MarkFox}, of the cavity mirrors to resolve the noninertial signal for various values of $\alpha$. The Unruh signal can be resolved from the inertial signal at lower accelerations inside a precisely designed cavity if the cavity mirrors have a correspondingly higher quality factor.}
    \begin{ruledtabular}
	    \begin{center}
		\begin{tabular}{c  c  c c}
        
			$ \alpha = a/\omega_0c $  & $R_{\rm min}$ & $\omega_0 L/c$= $2\pi L/\lambda_0$ & $ Q_{\rm min} $   \\
			\hline
			$10^{- 11}$  & $ 1 - 10^{-9} $ & $\pi- 10^{-8} $ & $\pi \times 10^{9}$\\
			
			$10^{- 10}$  & $ 1 - 10^{-8} $ & $\pi- 10^{-7} $ & $\pi \times 10^{8}$\\
			
			$10^{- 9}$  & $ 1 - 10^{-7} $ &  $\pi- 10^{-6} $ & $\pi \times 10^{7}$\\
			
		\end{tabular}
	\end{center}
    \end{ruledtabular}
	\end{table}
\paragraph*{Implementation\textemdash}
The requirement of acceleration and the recent demonstration of robustness of collective effects in relatively noisy conditions in NV centers in a diamond membrane coupled to a high-finesse cavity~\cite{Pallmann2024} suggest that such solid-state platforms can be used to test for the early superradiant burst caused by the Unruh effect. An atomic transition frequency of the order of $10 \,$GHz would be ideal as high quality factors $(Q \sim 10^9)$ have been reported for cavities having resonance frequencies in the range of few to tens of gigahertz. Moreover, microwave resonator oscillators with a fractional frequency stability $\sim 10^{- 16}$ have been demonstrated~\cite{Luiten1993,Creedon2010,Hartnett2012,Grop2014,Fluhr2023a,Fluhr2023b}, giving an uncertainty $\lesssim 10^{- 16}$ in the fractional length specification of the cavity. The idea of \textit{time-resolution accompanying superradiant amplification} can be first tested in compact prototype experiments subjecting a collection of atoms to non-linear accelerations~\cite{Lochan:2019,Biermann2020,Blencowe2021,Arya2022,Arya2023,Zheng2025,Zhou2025}, and potentially in analog systems~\cite{Alsing2005,Felicetti2015,Gooding2020,Tian2024,Katayama2025}. In all these scenarios, the modified stronger field fluctuations experienced by the accelerated sample should lead to an early superradiant burst under appropriate conditions. 
\paragraph*{Conclusion\textemdash}
We have addressed three key challenges facing any potential experimental enterprise to observe the Unruh effect. The requirement of extreme acceleration can be traded for high quality factor inside a precisely designed sub-resonant cavity. By identifying conditions under which a superradiant burst is exclusively seeded by the Unruh effect, we theoretically demonstrated a highly amplified and temporally resolved Unruh signal. A clear separation in time between the Unruh and the inertial signals addresses the challenge of inertial noise overwhelming the purely-noninertial signal.
We thus identify laboratory conditions and tools that magnify the subtle observer-dependent field theoretic effects to a detectable level. Finally, the equivalence principle points to the gravitational analogues of the effects presented here\textemdash gravity-induced spectral broadening of atoms, and a collective quantum response seeded by gravity~\cite{Arya2024b}.

\paragraph*{Acknowledgments\textemdash}
A.D. acknowledges funding from IISER Mohali. N.A. thanks Jorma Louko, Jerzy Paczos, and Magdalena Zych for helpful discussions and useful comments. 
N.A. acknowledges funding from Knut and Alice Wallenberg foundation through a Wallenberg Academy Fellowship No. 2021.0119.  Research of K.L. is partially supported by ANRF, Government of India, through a MATRICS research grant
no. MTR/2022/000900.
\paragraph*{Author Contributions\textemdash} N.A. and A.D. conceptualized the research. A.D. and N.A. performed the formal analysis and investigation. All the authors contributed to the interpretation and validation of the results and to the manuscript writing.
\paragraph*{Data availability\textemdash}
The data that support the findings of this article are not publicly available. The data are available from the authors upon reasonable request.
%
\onecolumngrid
\begin{center}
	\large \textbf{Supplemental Material: Time-resolved and Superradiantly Amplified Unruh Effect}
\end{center}

\renewcommand{\thesection}{S\arabic{section}}
\renewcommand{\theequation}{S\arabic{equation}}
\renewcommand{\thefigure}{S\arabic{figure}}
\section{Superradiant Master Equation}
Consider an ensemble of $N$ identical two-level atoms taking positions $\vb{r}_i$, coupled to a real massless quantum scalar field $\hat{\Phi}(\tau, \vb{r}_i)$ between two parallel mirrors with separation $L$, and complex reflection and transmission coefficients $r$ and $t$, respectively. Each atom carries a monopole moment $\hat{\mathfrak{m}}_j = i (\hat{\sigma}^{-}_j - \hat{\sigma}^{+}_j)$, where $\hat{\sigma}^{\pm}_j$ are the atomic raising and lowering operators for the $j$th atom. The free Hamiltonian of each atom is $\hat{H}_{j} = \omega_0 \hat{\sigma}^{z}_{j}/2$, where $\hat{\sigma}^{z}_{j}$ is the Pauli $z$-matrix for the $j$th atom.
The atom-field interaction Hamiltonian in the comoving frame of the atoms: $\hat{H}_{\rm I} = g \sum_{i=1}^{N} \hat{\mathfrak{m}}_i(\tau) \otimes \hat{\Phi}(\tau,\vb{r}_i)$, $g$ being the coupling constant, corresponds to the atom-light interaction in the dipole-approximation, simplified to describe interaction between an atom and a single polarization of the electromagnetic field~\cite{Soda2022,Martinez2013,KempfEduardo2014}. 

The Lindblad master equation governing the dynamics of the atomic array is given as~\cite{Gross1982}

\begin{equation}\label{ME}
	\dv{\hat{\rho}(\tau)}{\tau} =   -i[\hat{H}_{\rm eff}, \hat{\rho}(\tau)]
	+ \sum_{i,j}^{}  \gamma_{ij}  \Big(\hat{\sigma}_j^- \hat{\rho}(\tau) \hat{\sigma}_i^+ - \frac{1}{2}\acomm{{\hat{\sigma}_i^+ \hat{\sigma}_j^-}}{\hat{\rho}(\tau)}\Big)+\sum_{i,j}^{}  \chi_{ij}  \Big(\hat{\sigma}_j^+ \hat{\rho}(\tau) \hat{\sigma}_i^- - \frac{1}{2}\acomm{{\hat{\sigma}_i^- \hat{\sigma}_j^+}}{\hat{\rho}(\tau)}\Big) ,    
\end{equation}

where $\hat{\rho}(\tau)$ is the reduced atomic density operator and $\tau$ is the proper time.
The first term on the right hand side of the above equation drives Hamiltonian evolution of the system with~\cite{Gross1982,Breuer2002} 
\begin{equation}\label{eq:Heff}
	\hat{H}_{\rm eff} = \sum_{i,j}\Omega_{ij} \hat{\sigma}_i^+ \hat{\sigma}_j^-,
\end{equation}
where $\Omega_{ij} = \Im{\Gamma_{ij}(\omega_0) + \Gamma_{ij}(-\omega_0)}$, $\Gamma_{ij}(\omega) = g^{2}\int_{0}^{\infty} d\tau e^{i \omega \tau} \langle \hat{\Phi}(\tilde{x}_i(\tau)) \hat{\Phi}( \tilde{x}_j(0)) \rangle$, and $\tilde{x}_i(\tau)$ is the spacetime trajectory of the $i$th atom parameterized by its proper time $\tau$. The second term in Eq.~\eqref{ME} controls the dissipative dynamics of the atoms, where $\gamma_{ij}$ is the Fourier transform of the two-point field correlation function:
\begin{equation}\label{eq:gammaija}
	\gamma_{ij}(\omega) = g^{2}\int_{-\infty}^{\infty} d \tau e^{i \omega \tau} \langle \hat{\Phi}(\tilde{x}_i(\tau)) \hat{\Phi}( \tilde{x}_j(0)) \rangle.
\end{equation}
and 
\begin{equation}
	\chi_{ij}(\omega)= \gamma_{ij}(-\omega)=  g^{2}\int_{-\infty}^{\infty} d \tau e^{-i \omega \tau} \langle \hat{\Phi}(\tilde{x}_i(\tau)) \hat{\Phi}( \tilde{x}_j(0)) \rangle. 
\end{equation}
The $\gamma_{ij}$ and $\chi_{ij}$ account for the emission and absorption processes, respectively. The absorption events are absent for an inertial atom coupled to the vacuum state of the field and for a Rindler atom they are inconsequential at the accelerations of interest to this work, as they are exponentially suppressed with respect to the emission events, $\chi_{ij} = \exp{-2\pi \omega_0/a} \gamma_{ij}$. Therefore, for all practical purposes we can neglect the absorption events.
Doing so, the total emission rate of the array is obtained from Eq.~\eqref{ME} as:  
\begin{equation}\label{drate}
	\Gamma(\tau) = \sum_{i=1}^{N} \gamma_{ii} \expval{\hat{\sigma}_i^+ \hat{\sigma}_i^-}(\tau) + \sum_{i \neq j=1}^{N} \gamma_{ij} \expval{\hat{\sigma}_i^+ \hat{\sigma}_j^-}(\tau),
\end{equation}
where $\langle \hat{\sigma}_i^+ \hat{\sigma}_j^- \rangle(\tau) \equiv \Tr_{\rm A}{(\hat{\sigma}_i^+ \hat{\sigma}_j^- \hat{\rho}(\tau))}$, with $\Tr_{\rm A}(\cdot)$ denoting the trace over atoms. For $ i\ne j $, $\gamma_{ij}$ quantifies the influence of $i$th and $j$th atoms on each other's dynamics.
For a one-dimensional array with interatomic spacing much larger than the transition wavelength of each atom,  we have $\gamma_{ij} = \gamma_0 \delta_{ij}$, leading to an incoherent emission rate of $\Gamma(\tau) = N\gamma_0 e^{- \gamma_0 \tau}$, where $\gamma_0$ is the spontaneous emission rate of a single atom. 

In order to obtain the two-point correlation function $\langle \hat{\Phi}(\tilde{x}_i(\tau)) \hat{\Phi}( \tilde{x}_j(0)) \rangle$ of the field (required to compute the transition rates and the cooperation $\gamma_{ij}$), we need appropriate modes for quantization of the scalar field, which incorporate loss through boundaries in terms of reflection and transmission coefficients~\cite{Martini1991}. Note that we can evaluate the two-point correlator of the scalar field either by expanding the field in terms of the Rindler modes or by plugging the trajectory
\begin{equation}
	t(\tau) = a^{-1} \sinh(a \tau),~ z(\tau) = a^{-1} \cosh(a \tau)
\end{equation}
of the Rindler atoms in the scalar field expanded in terms of the inertial modes. We take the latter route. To this end, next, we obtain the field modes inside a lossy planar cavity.

\section{Field Quantization Between Two Leaky Parallel Mirrors}
Let us consider two infinite parallel mirrors P1 and P2 at $ x=-L/2 $ and $x = + L/2 $ in Cartesian coordinates, forming a planar cavity. Unitarity of operation requires the complex reflection $(r_1, r_2)$ and transmission $(t_1, t_2)$ coefficients of the mirrors to satisfy~\cite{Martini1991}: 
\begin{equation}
	\begin{split}
		\abs{r_1}^2 + \abs{t_1}^2=\abs{r_2}^2 + \abs{t_2}^2=1, \\
		r_1^*t_1 + r_1t_1^*=	r_2^*t_2 + r_2t_2^*=0.
	\end{split}
\end{equation}
\begin{figure}[h!]
	\includegraphics[width=0.4\linewidth]{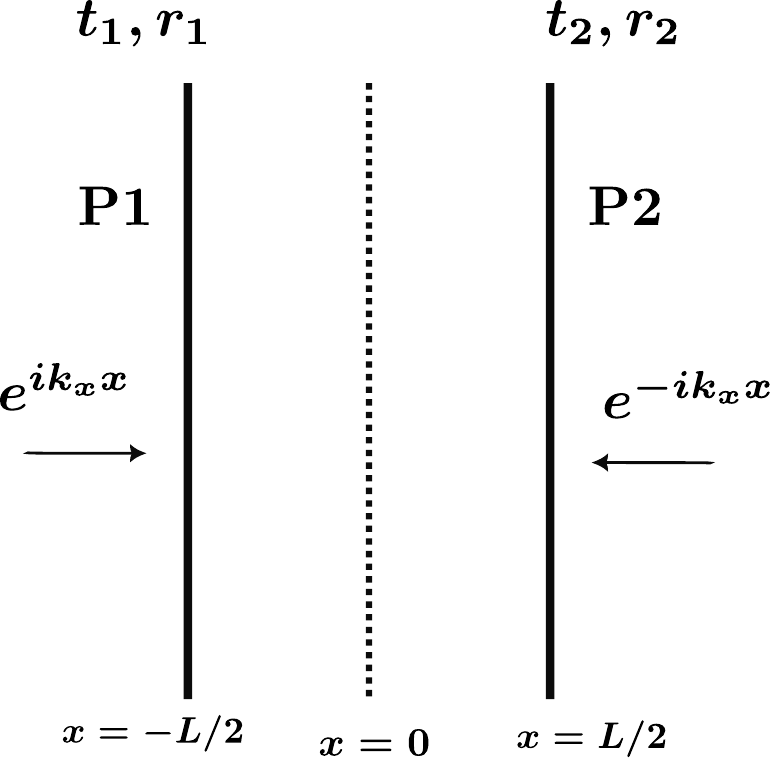}
	\caption{Cross-sectional view of the planar cavity formed by two parallel mirrors with complex transmission and reflection coefficients $t_i$ and $r_i$, respectively. The field modes between the mirrors are determined from the plane wave modes $e^{\pm i k_x x}$ incident on the mirrors~\cite{Martini1991}.}
\end{figure}
As we are interested in spatial modes between mirrors, note that plane waves in $\hat{x}$ direction incident on mirrors from left and right give rise to two distinct spatial mode profiles in between the mirrors depending upon reflection and transmission coefficients \cite{Ley1987}, 
\begin{align}
	U_{k_x}(x) &=	\frac{1}{D} \left( t_1 e^{i k_x x} + t_1 r_2 e^{-ik_xx + ik_x L} \right)~\text{and} \\
	U'_{k_x}(x) &=	\frac{1}{D} \left( t_2 e^{-i k_x x} + t_2 r_1 e^{ik_xx + ik_x L}\right),
\end{align}
where $D = 1- r_1 r_2 e^{2i k_x L}$ and $k^2 = k_x^2+k_y^2+k_z^2$.  The creation and annihilation operators $\hat{a}_k, \hat{a}_k^\dagger $ for $ U_{k_x} $, and $\hat{a}'_k, \hat{a}_k'^\dagger$ for $U'_{k_x}$ satisfy the commutation relations $[\hat{a}_k, \hat{a}_{k'}^\dagger] = [\hat{a}_k',\hat{a}_{k'}'^\dagger] = \delta(k-k')$, and $[\hat{a}_k,\hat{a}_{k'}'^\dagger] = [\hat{a}_k',\hat{a}_{k'}^\dagger]=0$.

Following the standard quantization procedure, the quantized scalar field between two parallel mirrors, providing for the mirror imperfections through non-ideal reflectivities, can be expressed as:
\begin{equation}\label{eq:qfield}
	\hat{\Phi}(\tilde{x})= \int \frac{d^3 k }{(2\pi)^{3/2}} \frac{1}{\sqrt{2 \omega_k}} \big(\hat{a}_k U_{k_x}(x) e^{- i \omega_k t + ik_y y + ik_z z} 
	+ \hat{a}_k' U'_{k_x}(x) e^{- i \omega_k t + ik_y y +ik_z z} + \text{h.c.} \big),
\end{equation}	
where $\tilde{x} \equiv (t,x,y,z)$ now denotes a spacetime event expressed in the inertial coordinates.
For symmetrical identical mirrors,  $ t_1 = t_2 = i\abs{t_1} = i\abs{t_2}=iT $ and $ r_1=r_2 = -\abs{r_1}= - \abs{r_2}=- R $, the two-point function of the field, $\mathcal{W}(\tilde{x}_i,\tilde{x}_j) = \langle \hat{\Phi}(\tilde{x}_i) \hat{\Phi}(\tilde{x}_j) \rangle $ can be computed to be:
\begin{multline}
	\mathcal{W}(\tilde{x}_i,\tilde{x}_j) =\int \frac{d^3k}{(2\pi)^3} \frac{1}{2 \omega_k} e^{- i \omega_k (t_i-t_j)}
	e^{i k_y \Delta y_{ij} + i k_z \Delta z_{ij}} \frac{2T^2}{\abs{D}^2}  \bigg\{(1+ R^2) \cos(k_x(x_i-x_j)) \\
	- 2R\cos(k_x (x_i+x_j))\cos(k_x L)\bigg\}.
\end{multline}
Note that we will be using  $\int d^3k = \int_{0}^{\infty}dk_{x} \int_{-\infty}^{\infty} dk_{y} \int_{-\infty}^{\infty}dk_{z}$ since the mode definitions at the beginning restrict the integration to one-half of the momentum space in the transverse direction to plates i.e. $x$ direction.
\subsection{$\gamma_{ij}$ for a Rindler Array}
\begin{figure}[h!]
	\centering
	\subfigure[]{\includegraphics[height=5cm]{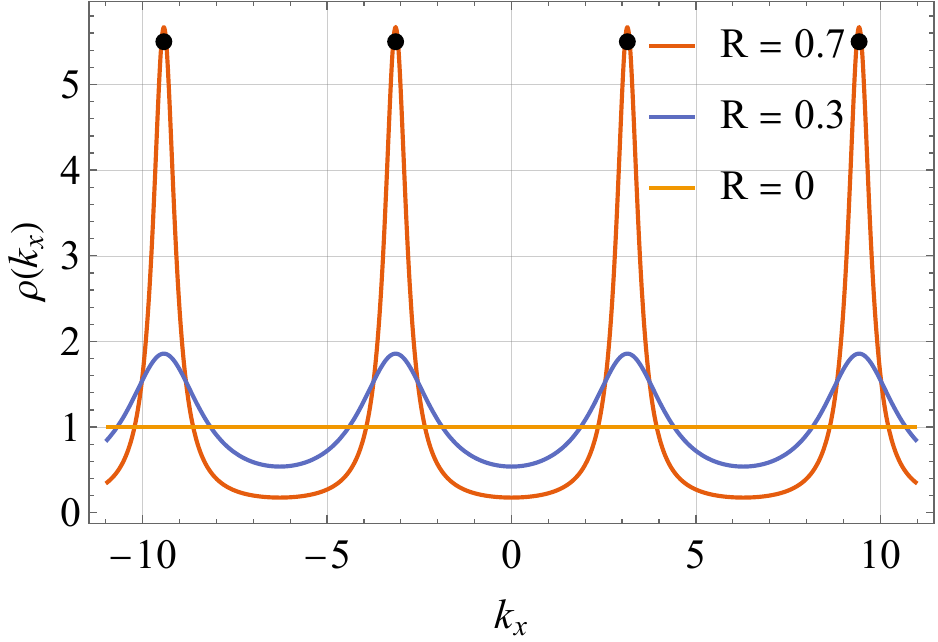}
		\label{fig:dos}}
	\subfigure[]{\includegraphics[height=5cm]{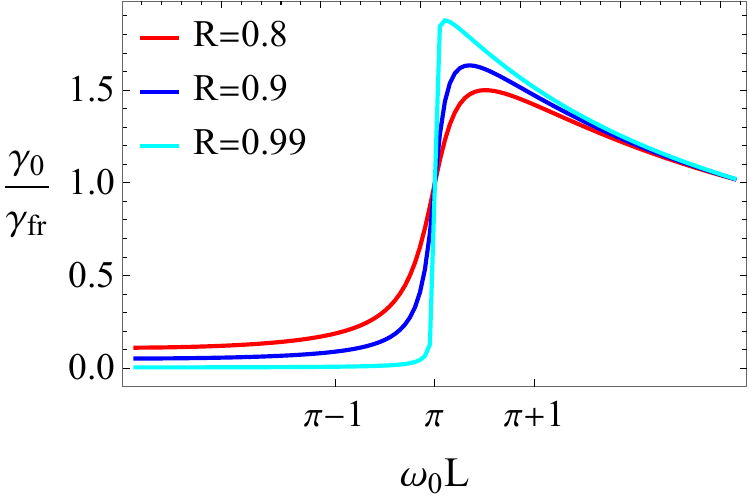}
		\label{fig:inertialg0lossy}}
	\caption{(a) The modification of the density of modes in the $x$ direction by the planar cavity for different values of the mirror reflectivity. As the mirror reflectivity (equivalently, the cavity quality factor) approaches zero, the free space is achieved with $\rho(k_x) = 1$ for all $k_x$. (b) Effect of cavity mirror reflectivity on spontaneous emission rate $\gamma_0$ of a single inertial atom contained in it as a function of the separation between the mirrors. Note that the first atom-cavity resonance point occurs for $\omega_0L=\pi $. To make the role of the cavity evident, the plot shows $\gamma_0$ normalized by the spontaneous emission rate $\gamma_{\rm fr}$ of the same atom in free space. For a lower mirror reflectivity, $\gamma_0$ is less suppressed in the sub-resonant cavity configuration.}
\end{figure}

Starting with Eq.~\eqref{eq:gammaija} and using field's expansion in terms of the inertial modes obtained in Eq.~\eqref{eq:qfield}, and plugging in the trajectory of a Rindler atom we obtain
	\begin{equation}
		\begin{split}
			\gamma_{ij}^{(\rm a)} =g^{2}\int_{-\infty}^{+\infty} \dd{s} e^{i \omega_0 s} \int \frac{d^2k_\perp}{(2\pi)^3}  \rho(k_x,L,R)e^{i k_y(y_i-y_j)}
			\int_{-\infty}^{+\infty} \frac{dk_z}{ \omega_k} \exp \left( -\frac{2i}{a}\sinh\left(\frac{as}{2}\right) \left[ \omega_k \cosh(a \tilde{\tau}) - k_z\sinh(a  \tilde{\tau}))\right]\right),
		\end{split}
	\end{equation}
where $s \equiv \tau_i - \tau_j$, $\tilde{\tau} \equiv (\tau_i + \tau_j)/2$, and
\begin{equation}\label{dos}
	\rho(k_x,L,R) \equiv \frac{(1+ R^2) - 2 R \cos(k_x L)}{(1- R^2) + \frac{4 R^2 \sin[2](k_x L)}{(1- R^2 )}},
\end{equation}
is the density of field modes as modified by the planar cavity formed by two parallel mirrors having reflectivity $R$. The density of field modes is illustrated in Fig.\,S\ref{fig:dos} for different values of the mirror reflectivity.

Transforming to new variables $\omega_k' = \omega_k \cosh(a \tilde{\tau}) - k_z\sinh(a \tilde{\tau})$ and $k_z'= k_z \cosh(a \tilde{\tau}) - \omega_k \sinh(a \tilde{\tau})$, one obtains 
\begin{equation}\label{gammaij} 
	\gamma^{(\rm a)}_{ij} =  g^{2} \int \frac{\dd^2 k_{\perp} }{(2\pi)^3} \int_{0}^{\infty} \dd \omega_k' \Theta(\omega_k' - k_{\perp}) \frac{\rho(k_x,L,R)}{\sqrt{\omega_k'^2 - k_{\perp}^2}}
	e^{i k_y \Delta y_{ij}} \int_{-\infty}^{\infty} \dd{s} e^{i\omega_0 s} \exp( - \frac{2i \omega_k'}{a} \sinh(\frac{as}{2})).
\end{equation}
Further, for $i=j$ and $a \to 0$, we obtain the spontaneous emission rate of an inertial atom inside a planar cavity. For higher values of the mirror reflectivity, the emission rate of an inertial atom in the sub-resonant cavity configuration is suppressed more strongly, as illustrated in Fig.~\ref{fig:inertialg0lossy}.
\section{Acceleration-induced Spectral Broadening}
Equation~\eqref{gammaij} can be cast in a more general form as $\gamma^{\rm (a)}_{ij} \propto \int_{0}^{\infty} \dd{\omega}_k' \rho(\omega_k') e^{i k_y \Delta y_{ij}} \mathcal{I}(\omega_k',\omega_0,a)$, where $\rho(\omega_k')$ is the density of field modes and $\mathcal{I}(\omega_k',\omega_0,a)$ decides the field modes participating in the atomic system's response.
The uniform acceleration induces non-resonant behavior in an atom, causing spectral broadening~\cite{Alsing2004,Jaffino2022}. The emission rate of a single inertial $(a = 0)$ and Rindler $(a \neq 0)$ atom can be obtained by setting $i=j$ in Eq.~\eqref{gammaij}.
The difference, $\gamma_{\rm a} - \gamma_{\rm 0}$, in the two responses hinges on the Rindler atom perceiving the Minkowski plane waves with a time-dependent phase as is evident in the $\exp(-2i (\omega_k'/a) \sinh(as/2))$ factor, as opposed to a factor of $\exp(- i \omega_k s)$ for an inertial atom (note that $\lim_{a \to 0} \exp(-2i(\omega_k'/a) \sinh(as/2)) = \exp(-i\omega_k s)$).

In the case of an inertial atom, $\mathcal{I}_{\rm inertial} \equiv \int_{-\infty}^{+\infty} \dd{s} e^{i\omega_0 s}  e^{- i \omega_k s} = 2 \pi \delta(\omega_k - \omega_0)$ enforces resonant coupling of the atom to field modes with frequency $\omega_0$.
In sharp contrast however, for a Rindler atom, the time integral doesn't lead to a Dirac delta function:
\begin{equation}
	\mathcal{I}_{\rm Rindler} \equiv \int_{-\infty}^{+\infty} \dd{s} e^{i\omega_0 s} e^{- 2i (\omega_k'/a) \sinh(as/2)}
	= \frac{4}{a} e^{\pi \omega_0/a} K_{2i \omega_0/a} \left(\frac{2\omega_0}{a} \frac{\omega'_k}{\omega_0}\right),
\end{equation}
where $K_{\nu}(x)$ is the modified Bessel function of the second kind. Note that $\mathcal{I}_{\rm Rindler}$ doesn't strictly enforce $\omega'_k/\omega_0 = 1$, unless $a \to 0$, as illustrated in Fig.\,2 of the main text. 
\section{Coherent dipole-dipole interactions}
The strength of the coherent dipole-dipole interactions in the atomic array is quantified by $\Omega_{ij}$~(see Eq.\eqref{eq:Heff}). For Rindler and inertial atomic arrays inside the planar cavity, we obtain 
\begin{equation}
	\Omega^{\rm (a)}_{ij}= - g^{2}\int \frac{d^2k_\perp}{(2\pi)^2} \rho(k_xL,R) e^{i k_y y_{ij}}  \frac{K_{i \omega_0/a}(k_\perp/a )}{a} \left(I_{- i \omega_0/a}(k_\perp/a ) + I_{i \omega_0/a}(k_\perp/a )\right),
\end{equation}	
and
\begin{equation}
	\Omega^{\rm (0)}_{ij} = - g^{2}\int \frac{d^2k_\perp}{(2\pi)^2} \rho(k_x L,R) e^{i k_y y_{ij}} \frac{\Theta(k_\perp -\omega_0)}{\sqrt{k_\perp^2-\omega_0^2}},
\end{equation}
respectively, where $I_{i \nu}(x)$ and $K_{i \nu}(x)$ are the modified Bessel functions of first and second kind, respectively. 

The field-mediated coherent dipole-dipole interactions between atoms in the array cause collective Lamb shift in each atom's energy levels~\cite{Friedberg1973,Gross1982}. The total Lamb shift (including both individual and collective contributions) of the atom at the $i$th site is given as $\Omega_i = \sum_{j=1}^{N} \Omega_{ij}$. The inhomogeneity in the array introduced by the coherent dipole-dipole interactions, that causes dephasing of the atomic dipoles, is given by the variance of $\Omega_i, i=1,2,\cdots, N$. The characteristic time scale over which dephasing occurs between a pair of atoms depends on the difference between the total phase acquired by the pair. To address the problem of dephasing, one effective approach involves configuring the spatial arrangement of atoms so that each atom experiences a symmetric environment. A symmetric environment ensures uniformity in the phase accumulation, thereby minimizing the disparities in the phase acquired by different atoms. A sufficiently long ordered array fulfills this requirement, as atoms in the bulk experience almost identical environment and accumulate phase uniformly, ensuring dephasing effects are minimal. 

\begin{figure}
	\includegraphics[scale=0.7]{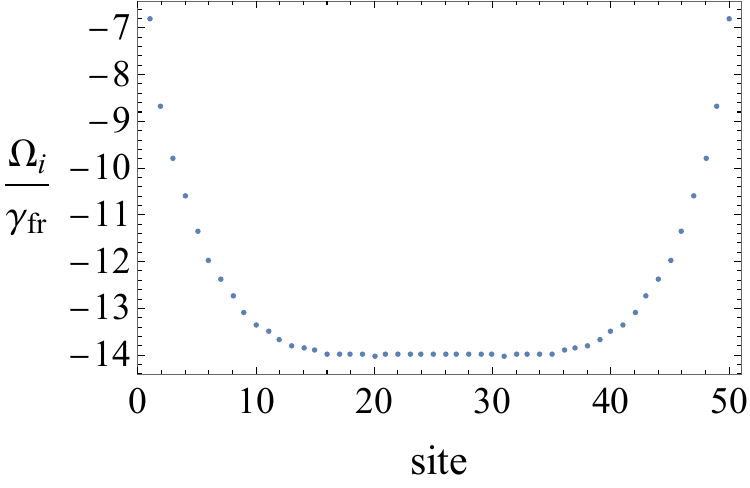}
	\caption{The cooperative Lamb shift as a function of site in a Rindler atomic array containing fifty atoms. Atoms in the bulk of the array experience almost identical environments and hence suffer similar shifts. Therefore, dephasing due to coherent dipole-dipole interactions is reduced in such ordered arrangements. The plot is for   $a/\omega_0 = 10^{-5}$ and $d/\lambda_0=1$.}
	\label{fig:depahsing50}
\end{figure}

For accelerations of interest, we confirm that this is also the case for a Rindler array. As shown in Fig.~\ref{fig:depahsing50}, the atoms in the bulk of an ordered array are shielded from this inhomogeneity as they experience an identical environment. Overall, for an appropriate choice of $d/\lambda_0$, the coherent dipole-dipole interactions between atoms cause only a slow dephasing even in an accelerated array. The atoms in the bulk thus remain indiscernible with respect to their coupling to the electromagnetic field, therefore, the evolution of these atoms proceeds in a permutationally invariant (under $i \leftrightarrow j$) manner.
\section{Total Emission Rate}
For the initial state of the array, we consider the product state
\begin{equation}
	\hat{\rho}(0) = \Pi_{i} \ket{\theta_{0},\varphi_{0}}_i  \leftindex_i {\bra{\theta_{0},\varphi_{0}}},~ \theta_{0} < \pi,
\end{equation}
where 
\begin{equation*}
	\ket{\theta_{0},\varphi_{0}}_i = \sin(\theta_0/2) e^{-i \varphi_0/2} \ket{e}_{i} + \cos(\theta_0/2) e^{i \varphi_0/2} \ket{g}_{i},
\end{equation*}
with $\ket{e}_i$ and $\ket{g}_i$ denoting the excited and ground states, respectively, of a two-level atom.

To obtain the total emission rate from Eq.~\eqref{drate}, we need $\expval{\sigma^{+}_{i} \sigma^{-}_{j}}(t)$. In general, the two-particle mean values depend on three-particle mean values and so on~\cite{Agarwal1970a,Agarwal1971c}. Thus, an exact solution of the problem requires solving an entire hierarchy of equations. Alternatively, one can resort to appropriate approximations to decouple the multi-particle mean values. Following~\cite{Agarwal1970a,Agarwal1971c}, we introduce the approximation $\expval{\sigma_i^z \sigma_j^z} = \expval{\sigma_i^z} \expval{\sigma_j^z}, \ \ (i\ne j) $ which introduces an inaccuracy of the order of $1/N$. Under this approximation, the total emission rate of the array is obtained as given in Eq.~(1) of the main text, as outlined below.

Note the identity
\begin{equation}
	\sum_{i \neq j} \sigma^{+}_i \sigma^{-}_j = \frac{1}{4} \sum_{i \neq j} (\va{\sigma}_i \cdot \va{\sigma}_j - \sigma^{z}_i \sigma^{z}_j),
\end{equation}
where $ \va{\sigma}_{i} \cdot \va{\sigma}_{j} = \sigma^{x}_{i}  \sigma^{x}_{j} + \sigma^{y}_{i}  \sigma^{y}_{j} + \sigma^{z}_{i}  \sigma^{z}_{j} $; and the fact that the permutation operator in terms of $\va{\sigma}$ is given by
\begin{equation}
	P_{ij} = \frac{1}{2} ( \mathds{1}+ \va{\sigma}_i \cdot \va{\sigma}_j).
\end{equation}
As noted in the previous section, for atoms in the bulk of a long ordered array the evolution remains permutationally invariant, therefore $\expval{\sigma_i^z}$ is the same for all such atoms and $\expval{P_{ij}} = 1$~\cite{Agarwal1970a,Agarwal1971c}. Under these observations, one obtains~\cite{Eberly1971,Agarwal1970a}
\begin{equation}
	\Gamma(\tau) = -\dv{W}{\tau} = \gamma_{0}\left[\mu \left(\frac{N^2}{4}  - W^2\right)   +  \left(\frac{N}{2} + W\right)\right],
\end{equation}
where $W= \sum_{k} \expval{\sigma_{k}^z}/2$ is the total (dimensionless) energy of the atomic system and we have defined $\mu = (\gamma_{0} N^2)^{-1} \sum_{i\ne j} \gamma_{ij}$.
Solving the differential equation with the initial condition $W(0) = - N \cos\theta_0/2$, we obtain
	\begin{equation}
		\Gamma(\tau) = \frac{\gamma_0 (1-\cos\theta_0) N (\mu  N+1)^2 [(1 + \cos\theta_0) \mu  N + 2] e^{\gamma_0 \tau (\mu  N+1)}}{\left([ \mu N (1 + \cos\theta_0 )+2] e^{\gamma_0 \tau (\mu  N+1)}+(1 - \cos\theta_0) \mu  N\right)^2}.
	\end{equation}
For $\mu>0$ and $\theta_0 \to \pi$, the above equation can be rearranged to obtain Eq.~(1) of the main text.

For a Rindler array, in principle the absorption events would also be present and therefore, staring with Eq.~\eqref{ME}, the total photon emission rate in the comoving frame of atoms is obtained as:
\begin{equation}\label{eq:RindlerEmRateFull}
	\Gamma_{\rm a}(\tau)=  \sum_{i,j} \gamma^{\rm (a)}_{ij} \expval{\sigma^{+}_{i} \sigma^{-}_{j}}(\tau) - \sum_{i,j} \chi^{\rm (a)}_{ij} \expval{\sigma^{-}_{i} \sigma^{+}_{j}}(\tau). 
\end{equation}
Terms with $\chi^{\rm (a)}_{ij}$ corresponds to the absorption terms due to acceleration. Under the same approximations as mentioned earlier in this section, $\Gamma_{\rm a}(\tau)$ can be cast as a differential equation for $W(\tau)$: 
\begin{equation}\label{eq:full-Gamma}
	\Gamma_{\rm a}(\tau) =  - \dv{W(\tau)}{\tau} =   \gamma_{\rm a}\left[\mu_{\rm a} \left(\frac{N^2}{4}  - W^2\right)   +  \left(\frac{N}{2} + W\right)\right] -\chi_{\rm a}\left[\mu_{\rm a} \left(\frac{N^2}{4}  - W^2\right)   +  \left(\frac{N}{2} - W\right)\right],
\end{equation}
here $\gamma_{\rm a} \equiv \gamma_{ii}^{\rm (a)}$ and $\chi_{\rm a} \equiv \chi_{ii}^{\rm (a)} = e^{-2\pi \omega_0/a} \gamma_{\rm a}$. Solving the above equation, with initial condition $W(\tau =0)= -N \cos \theta_{0}/2$ in comoving frame of the array, one obtains:
\begin{equation}\label{eq:full-Gamma2}
	\Gamma_{\rm a}(\tau) = -\frac{N\varpi  \left[4(\chi_{\rm a}+\gamma_{\rm a})\cos \theta_{0} +(\chi_{\rm a}-\gamma_{\rm a})(4+\mu_{\rm a} N - \mu_{\rm a} N\cos2\theta_{0})\right]e^{\tau \sqrt{\varpi}}}{2\left[-(\chi_{\rm a}+\gamma_{\rm a}-\mu_{\rm a}  N (\chi_{\rm a}-\gamma_{\rm a}) \cos \theta_{0} )+e^{\tau \sqrt{\varpi}} \left(\chi_{\rm a}+\gamma_{\rm a}-\mu_{\rm a}  N (\chi_{\rm a}-\gamma_{\rm a}) \cos \theta_{0} +\sqrt{\varpi}\right)+\sqrt{\varpi}\right]^{2}},
\end{equation} 
where $\varpi \equiv 4 \chi_{\rm a} \gamma_{\rm a}+(\chi_{\rm a}-\gamma_{\rm a})^2 (\mu_{\rm a}  N+1)^2$.
However, as noted earlier the absorption events are highly suppressed compared to the emission events in the low-acceleration $(a/\omega_0 \ll 1)$ regime of interest. Therefore, they can safely neglected.
Further, after neglecting the absorption events, Eq.~\eqref{eq:full-Gamma2} can be cast in the form of Eq.~(1) of the main text for $\theta_0 \to \pi$.
\section{Cavity's Finite Quality Factor, Imperfect Atomic Cooperation, and Time-resolution of the Unruh Signal}

To employ the time-resolving nature of the superradiance process one needs to account for the factors to which $\tau_{\rm d}$ and $\tau_{\rm sr}$ are intrinsically sensitive, these include (i) cavity's quality factor $(Q)$, (ii) separation between the atoms, and (iii) dephasing due to coherent dipole-dipole interactions.
In this section, we discuss how an incomplete inclusion of the superradiance process, simply as a perfect amplifier, can lead to severe ambiguities in identifying the cause of the signal obtained in an experiment set up to observe the Unruh or the related effects. We illustrate this below with explicit examples.

\subsection{Trade-off between acceleration and cavity's quality factor} 
Consider the trade-off between the acceleration and minimum cavity quality factor required to resolve the corresponding Unruh signal, as summarized in Table I of the manuscript, and its consequences for the cause of the superradiant burst.
\begin{figure}[h!]
	\centering
	\includegraphics[width=0.7\linewidth]{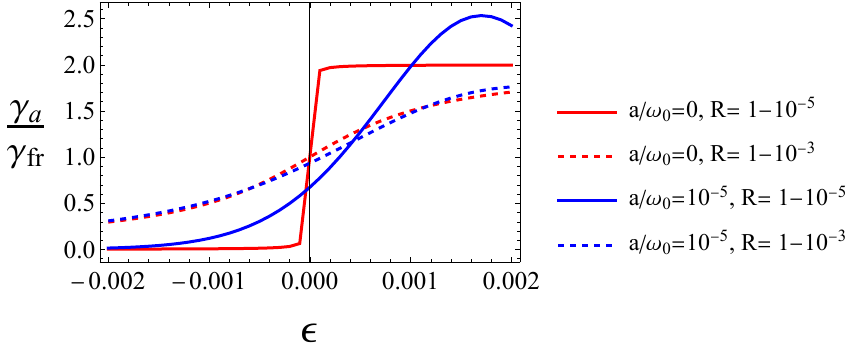}
	\caption{Trade-off between acceleration and the minimum cavity quality factor required to resolve the Unruh signal.}
	\label{fig:a-and-Q-tradeoff}
\end{figure}
Fig.~\ref{fig:a-and-Q-tradeoff} demonstrates an example of the tradeoff. For $\alpha \equiv a/\omega_0 c = 10^{-5}$, the inertial $(\gamma_0/\gamma_{\rm fr})$ and the noninertial $(\gamma_{\rm a}/\gamma_{\rm fr})$ signals are:
{\renewcommand{\labelenumi}{(\Alph{enumi})}
	\begin{enumerate}
		\item well-resolved (solid curves) for $Q = \pi \times 10^5$, with the acceleration-induced correction, $\tilde{\gamma}(\alpha) \equiv \gamma_{\rm a} - \gamma_0$, to the emission rate of a single atom much greater than the emission rate $\gamma_0$ of an inertial atom in the sub-resonant cavity configuration.
		
		\item not resolved (dashed curves) for $Q = \pi \times 10^3$ as they effectively overlap, meaning that the acceleration-induced correction $\tilde{\gamma}(\alpha)$ is negligible.  
\end{enumerate}}
If we employ superradiance in these configurations of the setup, we get an \textit{early} superradiant burst exclusively seeded by the Unruh effect in configuration (A). That is, if we analyze both an inertial and a Rindler sample of atoms in configuration (A), we will obtain a superradiant burst in each case, with the burst for Rindler sample occurring considerably earlier.  

However, in configuration (B) of the setup, the bursts in the two cases will overlap in time because both the bursts are effectively seeded by the inertially-perceived field fluctuations. An analysis assuming a perfect cavity ($Q \to \infty$) would be oblivious of the acceleration-quality factor tradeoff reported here~\cite{Zheng2025}. Such an analysis would expect only a single burst and therefore the overlapping bursts in configuration (B) would report a false detection of the Unruh effect. The time-resolving approach introduced in this work accounts for a finite $Q$ and therefore can tell the configurations (A) and (B) apart. This example also contrasts the present scheme from other schemes that employ cavity-QED systems for detecting the Unruh effect at low accelerations based only on intensity discrimination~\cite{Jaffino2022,Arya2024ULS,Barman2025}. An integrated approach relying both on intensity and temporal discrimination provides a more robust mechanism for detecting a signal that is intrinsically extremely weak.

\subsection{Imperfect Cooperation Among Atoms}
\begin{figure}[h!]
	\centering
	\subfigure[]{\includegraphics[width=0.45\linewidth]{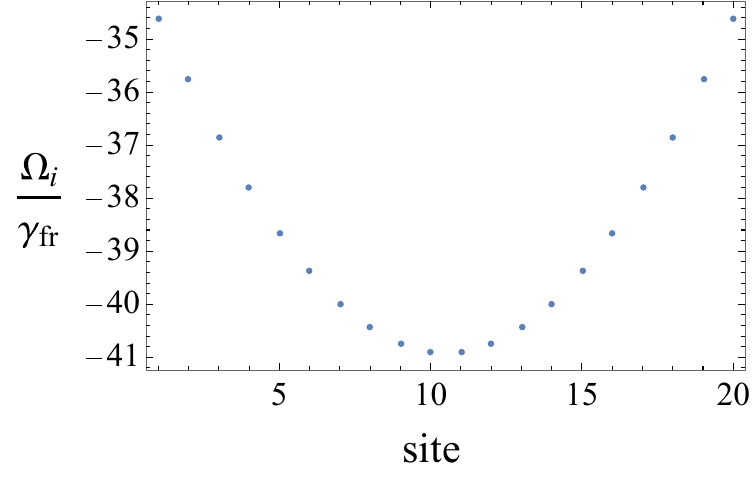}
		\label{fig:dephasing0.1}}
	\subfigure[]{\includegraphics[width=0.45\linewidth]{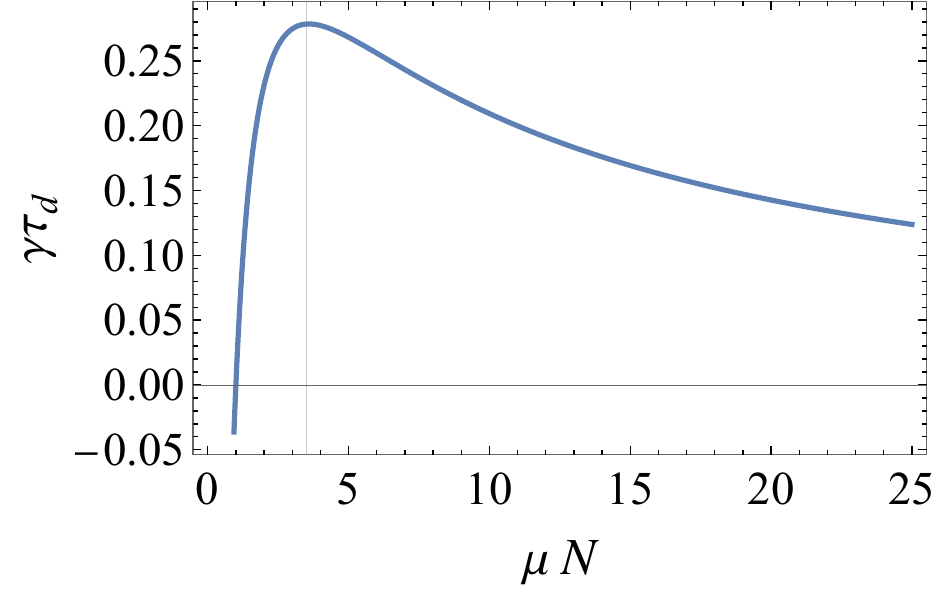}
		\label{fig:td-vs-muN}}
	\caption{(a) Greater inhomogeniety in collective Lamb shift along the atom array for a smaller interatomic spacing of $d=0.1 \lambda_0$. Compare with figure \ref{fig:depahsing50}. (b) Variation of the superradiant delay time with the effective number of cooperating atoms $\mu N$. }
	\label{fig:cooperation}
\end{figure}

The separation between the atoms plays a crucial role in determining the timing of the superradiant burst as it affects the effective number of cooperating atoms $\mu N$. Figure~\ref{fig:cooperation}(a) shows collective Lamb shift in each atom in a one-dimensional array of $20$ atoms, with interatomic spacing $d = 0.1 \lambda_0$, acceleration $a/\omega_0 c = 10^{-5}$, and cavity quality factor $Q = \pi \times 10^5$.
The figure shows that the collective Lamb shift in each atom is very inhomogeneous across the array. As a result, the atoms don't remain indistinguishable with respect to their coupling to the cavity field, and, consequently, the effective number of atoms cooperating is very low. Figure~\ref{fig:cooperation}(b) shows how $\gamma \tau_{\rm d}$ varies with $\mu N$. The timing of the superradiant burst can shift earlier or later (including the possibility that a burst is not observed at all), at a fixed acceleration of the array, depending on the factors influencing $\mu N$ like the dephasing caused by the coherent dipole-dipole interactions and cavity detuning. Therefore, one needs to benchmark the actual number of atoms cooperating in the setup as we have done in Fig.~4(c) of the main text by plotting the ratio of $\mu_{\rm a}/\mu_{0}$ (for $a/\omega_0 c = 10^{-9}$ and $d = \lambda_0$), as a function of the cavity detuning. Fig.~4(c) of the main text shows that under these conditions, $\mu N$ will be the same for both an inertial and Rindler array. Thus, the expected time of the superradiant burst seeded by the Unruh effect can be determined from Fig.~\ref{fig:cooperation}(b) by setting $\gamma = \gamma_{\rm a}$.

In summary, incomplete treatments of superradiance where it is incorporated simply as a perfect amplifier, inevitably introduce severe ambiguities regarding the cause of the superradiant burst. A burst seeded by the inertially perceived vacuum fluctuations can very easily be mistaken for the one caused by the Unruh effect.
\section{Considerations on a Potential Implementation}

\textit{Arrangement of atoms and the required size of the sub-resonant cavity:} We have considered a one-dimensional array with an interatomic spacing of the order of atomic transition wavelength to mitigate the dephasing effects due to coherent dipole-dipole interactions between the atoms. Mitigating this noise requires that all atoms in the sample experience an almost identical environment so that the coherent dipole-dipole interactions do not lead to relative phase between the atoms. More precisely, the dephasing time scale must be longer than the superradiance delay time~\cite{Friedberg1974,Coffey1978,Gross1982}. To keep the required cavity size small, one can employ atoms arranged in a ring configuration. The ring configuration provides identical environment for all the atoms and has been shown to be immune to the above-mentioned dephasing~\cite{Coffey1978,Gross1982}.

We envision about 15 two-level systems arranged in a ring configuration. This will potentially require NV centers on a diamond substrate to support acceleration. We note that collective effects in NV centers in diamond have been recently experimentally observed under much more noisy conditions~\cite{Pallmann2024} than we have allowed in our calculations. More generally, an exact spatial placement of atoms may not be necessary. As exemplified by ref.\,\cite{Pallmann2024}, all one requires is a platform capable of allowing collective effects in quantum emitters, that is, any dephasing must be slow enough such that the emitters can acquire the quantum coherence required for a superradiant burst. Once bench-marked with superradiance from an inertial sample, such a setup would serve our purpose well.

\textit{Required cavity mirror quality factor and precision in cavity size:} An atomic transition frequency of the order of $10 \,$GHz would be ideal for our proposal as high quality factors $(Q \sim 10^9)$ have been reported for cavities having resonance frequencies in the range $\sim 1-10 \,$GHz. Moreover, microwave resonator oscillators with a fractional frequency stability $\sim 10^{- 16}$ have been demonstrated~\cite{Luiten1993,Creedon2010,Hartnett2012,Grop2014,Fluhr2023a,Fluhr2023b}. That gives an uncertainty $\sim 10^{-6}$ in the length specification of a cavity with resonant frequency $\sim 10 \,$GHz. Inside a cavity with quality factor $\sim 10^9$, a sample of two-level systems, each with a resonant frequency $\sim 10\,$GHz, would require an acceleration $\sim 10^7 \,\mathrm{m/s^2}$ to cause an early superradiant burst seeded by the Unruh effect.

In addition, we also point out several platforms capable of realizing the \textit{analog} Unruh effect. For example, collective effects in up to ten superconducting qubits have been observed~\cite{Mlynek2014,ZhenWang2020} and an effective linear acceleration $\sim 10^{15}\,\mathrm{m/s^2}$ can be emulated in a superconducting cavity-QED setup by modulating qubit-field coupling in time~\cite{Felicetti2015}. We therefore expect rapid progress towards first experimental observations of the analog Unruh phenomena in these platforms. Such proof-of-principle experiments would serve to confirm the time-resolved and superradiantly amplified \textit{analog} Unruh effect and would precipitate efforts towards direct observations.
\end{document}